\documentclass[%
 reprint,
 amsmath,amssymb,
 aps,
 nofootinbib,
]{revtex4-2}

\usepackage{mathrsfs}
\usepackage{graphicx}
\usepackage{dcolumn}
\usepackage{bm}

\begin{document}

\preprint{APS/123-QED}

\title{Symplectic structure of equilibrium thermodynamics}

\author{Luis Arag\'on-Muñoz}
\affiliation{Instituto de Ciencias Nucleares, Universidad Nacional Aut\'onoma de M\'exico, 
	04510 Ciudad de M\'exico, M\'exico }
 \email{luis.aragon@correo.nucleares.unam.mx}
\author{Hernando Quevedo}%
 \email{quevedo@nucleares.unam.mx}
\affiliation{Instituto de Ciencias Nucleares, Universidad Nacional Aut\'onoma de M\'exico, 
	04510 Ciudad de M\'exico, M\'exico }
\affiliation{Dipartimento di Fisica and ICRA,  Universit\`a di Roma ``La Sapienza", I-00185 Roma, Italy}

\date{\today}

\begin{abstract}
The contact geometric structure of the thermodynamic phase space is used to introduce a novel symplectic structure on the tangent bundle of the equilibrium space. Moreover, it turns out that the equilibrium space can be interpreted as a Lagrange submanifold of  the corresponding tangent bundle,  if the fundamental equation is known explicitly. As a consequence, Hamiltonians can be defined that describe thermodynamic processes.
\begin{description}
\item[Keywords]
Contact geometry, symplectic geometry, thermodynamics, Hamiltonian systems
\item[MCS numbers]
53D05,53D10,80A05
\end{description}
\end{abstract}

\maketitle


\section{Introduction} 

Different geometric methods have been applied for a long time in thermodynamics to formulate in an alternative manner its fundamental laws and conceptual background. 
From a geometric point of view, thermodynamic systems and their quasi-static processes can be described by using contact geometry \cite{Mrugala,Zorich,Arkady}. In particular, one can introduce a contact manifold called { thermodynamic phase space}, which for each thermodynamic system contains an embedded manifold called equilibrium space, consisting of the topological collection of all equilibrium states represented as points of the space. 
This allows us, for instance, to understand different thermodynamic representations as expressions of the same theory in different coordinates. Moreover, a differential contact form for the thermodynamic first law appears naturally by demanding that the equilibrium space be a submanifold, where its contact form is zero, called  Legendre submanifold.

In this work, we focus on the contact structure of the phase space and the differential structure of the equilibrium space to investigate the possibility of introducing a compatible symplectic structure. We will see that this is possible if we consider the tangent bundle of the equilibrium space. Indeed, it turns out that a procedure called contactization can be applied to this tangent bundle and the resulting manifold can be identified with the phase space with its contact structure. As a byproduct of this procedure, the Hamiltonian description is obtained canonically. Moreover, the Hamilton-Jacobi conceptual basis can also be obtained as a result of the contactization and the analysis of the corresponding Hamilton-Jacobi equation shows that its solutions determine complete families of thermodynamic systems, which are otherwise not related at all. 

This work is organized as follows. In Sec. \ref{sec:level2}, we review the main concepts that are used for formulating a geometric description of equilibrium thermodynamics. In particular, the equilibrium space and the phase space are defined with the level of rigor that is necessary to investigate the underlying symplectic structure. In Sec. \ref{sec:bundle}, we interpret the thermodynamic phase space as a principal fiber bundle and the contact distribution as a horizontal distribution. By calculating the holonomy in horizontal lifted curves, we find an indication of the existence of an underlying symplectic structure. Then, the contactization of the cotangent bundle of the equilibrium space is performed in Sec. \ref{sec:contact}. The Hamiltonian and Hamilton-Jacobi structures that arise naturally as a result of the contactization are also investigated and applied to particular examples. In Sec. \ref{sec:group}, we extend the notion of Legendre's total transform to the framework of symplectic geometry and analyze the implications. 
In particular, we  find a momentum map associated with the action of the maximal torus of the rotation group in $\mathbb{R}^{2n}$ over the cotangent bundle of the equilibrium space. Finally, in Sec. \ref{sec:con}, we discuss our results and comment on possible applications for future works.

\section{\label{sec:level2}Thermodynamics}

Let $\mathrm{A}$ be a thermodynamic system, i.e., a portion of the universe with a set $\{a,b,c,...\}$ of special states, called states of equilibrium, in which $\mathrm{A}$ is described by a small number $n\in \mathbb{N}$ of bulk measurements $(q^{1},q^{2},...,q^{n})$, called thermodynamic variables, whose values are given according to $n$ equations of the form $(i=1,2,...,n)$
\begin{equation}
\label{eq:1}
p_{i}\,=\,p_{i}\left(q^{1},...,q^{n}\right)\,\equiv\,\displaystyle\frac{\partial \phi}{\partial q^{i} }.
\end{equation}
In Eq.(\ref{eq:1}), the functions $p_{i}$ are also known as  thermodynamic variables\footnote{In particular, when the variables $(q^{1},...,q^{n})$ are all extensive, that is, they scale with the size of the system, the dual variables $p_{i}$ are also known as intensive variables, that is, zero-degree homogeneous functions of the $q'$s \cite{Callen}.}, that are specifically identified as the dual variables to $q^{i}$ and  represent control parameters, which are used to determine when two systems that interact and exchange matter would reach the equilibrium as a coupled system \cite{Callen}. The quantity $\phi$ is known as the thermodynamic potential and its relationship with the thermodynamic variables in the form of a function $\phi:\mathbb{R}^{n}\rightarrow \mathbb{R}$ is the so-called fundamental equation of the system \cite{Callen}. This election of the terms (dual variable, thermodynamic potential, and fundamental equation) originates from the differential of the function $\phi$ 
\begin{equation}
\label{eq:2}
d\phi\,=\,\displaystyle\sum\limits_{i=1}^{n}\displaystyle\frac{\partial \phi}{\partial q^{i}}dq^{i}\,=\,\displaystyle\sum\limits_{i=1}^{n}p_{i}dq^{i}.
\end{equation}
This equation can be interpreted as a conservation law  made up by the amount of terms $p_{i}dq^{i}$, which are formed by the product of the differential of a thermodynamic variable and its dual, representing  infinitesimal flows of matter in the system \cite{Callen}. The behavior of the function $\phi$, taking into account certain restrictions on the thermodynamic variables, determines the final equilibrium state of $\mathrm{A}$ \cite{Callen}. For example, for a simple thermodynamic system with thermodynamic variables $(q^{1},q^{2},q^{3})=(U,V,N)$, where $U$ is the energy, $V$ is the volume, and $N$ is the number of particles in the system, the entropy $S$ represents a thermodynamic potential, $S=S(U,V,N)$, which determines the final state of the system, maximizing the function $S=S(U_{0},V,N)$ for $U_{0}=\textup{const}$.

The description of a thermodynamic system in terms of $n$ real numbers, i.e., the thermodynamic variables $(q^{1},...,q^{n})$, and a potential $\phi=\phi(q^{1},...,q^{n})$ is not unique in the sense that it is always possible to interchange any coordinate $q^{i}$ with its dual coordinate $p_{i}$, maintaining  an extremal principle on the function \cite{Callen},

\begin{equation}
\label{eq:3}
\phi[p_{j}]\equiv\phi-p_{j}q^{j},
\end{equation}

\noindent where $q^{j}=q^{j}(q^{1},...,q^{j-1},p_{j},q^{j+1},...,q^{n})$. For this interchange to be possible, it is sufficient and necessary that $\phi$ be a convex function \cite{Callen, Arnold, Cahill}, a condition that from now on we will assume to be always valid for every thermodynamic potential.

Formally, Eq.(\ref{eq:3}) represents a partial Legendre transformation of the convex function $\phi=\phi(q^{1},...,q^{n})$ of $n$ real variables with respect to its	coordinates $q^{i}$ \cite{Callen}. Its generalization can be defined as follows: Let  $(I,J)$ be any partition of the index set $\{1,2,...,n\}$, i.e., $I$ and $J$ are  disjoint sets such that $I\cup J=\{1,2,...,n\}$. The Legendre transformation of the potential $\phi$ with respect to the interchange of the thermodynamic variables $q^{i}$ with their dual variables $p_{i}$ is given by

\begin{equation}
\label{eq:4}
\phi[p_{I}]\equiv \phi(p_{I},q^{J})\,-\,\displaystyle\sum\limits_{k\,\in\,I}p_{k}\,q^{k},
\end{equation}

\noindent with $p_{I}=\{p_{k}\}_{k\in I}$ and $q^{J}=\{q^{k}\}_{k\in J}$. Identifying the change of variables $(q^{I},q^{J})\rightarrow (q^{'I}=p_{I},q^{'J}=p_{J})$  given by the Legendre transformation (\ref{eq:4}), it is easy to see that the equations of state are modified according to the rule

\begin{equation}
\label{eq:5}
p'_{i}\equiv \displaystyle\frac{\partial \phi[p_{I}]}{\partial q^{'i}}\,=\,\left\{\begin{array}{lll}
\displaystyle\frac{\partial \phi}{\partial q^{i}}\,=\,p_{i},&\qquad&i\,\in\,J\\
\\
-\displaystyle\sum\limits_{k\in I}p_{k}\delta^{k}_{i}=-p_{i},&& i\,\in\,I.
\end{array}\right.
\end{equation}

In this sense, under coordinates changes that interchange variables with its duals, the original thermodynamic potential loses its nature as potential in the new coordinates \cite{Callen}. This leads us to think that thermodynamics as a mathematical theory is not totally free of coordinates; hence it has to be formulated in terms of a set of equilibrium states $\{a,b,c,...\}$ and an specific election of thermodynamic variables and potential: 
\begin{equation}
\label{eq:6}
\left[\!\!\begin{array}{c}
\textup{Thermodynamic system}
\end{array}\!\!\right]\!\!=\!\!\left(\{a,b,c,...\},\phi(q)\right).
\end{equation}
Since there is a total of $2n$ different choices of thermodynamic variables and potentials, there exist $2n$ Legendre transformations  that can be applied to describe only one system with $n$ degrees of freedom. We say then that thermodynamics presents a discrete symmetry under Legendre transformations. Nonetheless, it is possible to find a geometric formulation of thermodynamics by using contact geometry and, 
as we will show in this work, symplectic geometry.

\subsection{Thermodynamic equilibrium space}

For any thermodynamic system $\mathrm{A}$, with $n$ thermodynamic degrees of  freedom, we define the \textit{thermodynamic equilibrium space} (TES) $\mathcal{E}$ as a smooth $n$-dimensional manifold, obtained by giving to the set of equilibrium states of the system a topology $\tau(\mathcal{E})$ and a differential structure \cite{Mrugala}. In this manifold, we can introduce a preliminary atlas composed by charts where the coordinates are identified with the usual thermodynamic variables of the system like energy $U$, volume $V$, number of particles $N$, magnetization $M$, etc., which we will identify particularly as

\begin{equation}
\label{eq:7}
(q^{1},...,q^{n})\,\stackrel{*}{=}\,(E^{1},...,E^{n}),
\end{equation}

\noindent where the asterisk represents the identification, which has the advantage of having a direct physical interpretation, but that in no way results fundamental. Accordingly, we can generalize the concept of thermodynamic potential as any smooth mapping from $\mathcal{E}$ to $\mathbb{R}$, $\phi \in \mathcal{C}^{\infty}(\mathcal{E})$, whose  coordinated restriction given in (\ref{eq:7}) leads to a convex function. In this sense, the dual coordinates $I_{i}$ are the components of the 1-forms $d\phi \in \Lambda^{1}(\mathcal{E})$ with the cotangent basis $(dE^{1},..,dE^{n})$ induced by the corresponding coordinates so that the equations of state are determined by the relationships

\begin{equation}
\label{eq:8}
d\phi\,=\,\displaystyle\sum\limits_{i=1}^{n}I_{i}dE^{i}\quad\Longrightarrow\quad 
I_{i}\,\stackrel{*}{=}\,\displaystyle\frac{\partial \phi}{\partial E^{i}}.
\end{equation}

As a particular example consider an ideal gas  enclosed in a region of a variable volume with a constant number of particles. This system has only two degrees of freedom so that the corresponding $\mathcal{E}$ becomes a two-dimensional smooth manifold. In this case, the system of physical coordinates (\ref{eq:7}) can be chosen as the pair $(E^{1},E^{2})=(U,V)$ and the thermodynamic potential as the entropy $S$, with the fundamental equation \cite{Callen}

\begin{equation}
\label{eq:9}
S(U,V)\,=\,N\,k_{B}\,\ln\left(U^{C_{V}}V\right),
\end{equation}

\noindent where $k_{B}$ represents Boltzmann's constant and $C_{V}$ is the specific heat at constant volume. In this case, the dual thermodynamic variable to $U$ is identified as the inverse of the temperature $T$, while the dual to $V$ is the combination $P/T$, with $P$ as the pressure of the system. Moreover, the equations of state are 

\begin{equation}
\label{eq:10}
I_{1}=\displaystyle\frac{1}{T}=\displaystyle\frac{N k_{B} C_{V}}{U},\qquad
I_{2}=\displaystyle\frac{P}{T}=\displaystyle\frac{N k_{B}}{V}.
\end{equation}

\subsection{Thermodynamic phase space}

The equilibrium space $\mathcal{E}$ can be considered as an $n$-dimensional submanifold of a larger $2n+1$-dimensional manifold called \textit{thermodynamic phase space} (TPS) $\mathcal{B}$ that is constructed as the space of all the variables of systems with $n$ degrees of freedom, their duals and the potential $(q^{1},...,q^{n},p_{1},...,p_{n},\phi)$, removing any possible relations between them so that they can be used as coordinates of a $2n+1$-dimensional smooth manifold \cite{Mrugala,Mrugala2,Quevedo}.

We assume that  $\mathcal{B}$ is a contact manifold, i.e., there  exists a codimension-one distribution that is co-orientable and maximally non-integrable \cite{Geiges,Etnyre}. In other words, there exists a smooth family of codimension-one subspaces of tangent spaces $T\mathcal{B}$, $\Pi: \mathcal{B}\rightarrow T\mathcal{B}$, that can be seen as the kernel $\Pi= \textup{ker}(\alpha)$ of the elements of a conformal family\footnote{If $\beta\in \Lambda^{1}(\mathcal{B})$ is an element of $[\alpha]$, there exists a non-zero function $f:\mathcal{B}\rightarrow \mathbb{R}\backslash\{0\}$ such that $\beta=f\,\alpha$.} $[\alpha]$ of 1-forms $\alpha \in \Lambda^{1}(\mathcal{B})$, called  contact 1-forms,  satisfying the condition of being maximally non-integrable \cite{Geiges,Etnyre,Blair}
\begin{equation}
\label{eq:11}
\alpha\,\wedge\,(d\alpha)^{\wedge n}\,\equiv\,\alpha\,\wedge\,\underbrace{d\alpha\wedge...\wedge d\alpha}_{n-\textup{times}}\,\neq\,0.
\end{equation}

This condition essentially means that there are maximally integral submanifolds $L$ tangent to the distribution $\Pi$ that can be interpreted as $n-$dimensional equilibrium spaces  \cite{Mrugala,Zorich,Arkady,Quevedo}. In general terms, the highest dimension that can have a submanifold $L$ is $\textup{dim}(L)=\frac{1}{2}(\textup{dim}(\mathcal{B})-1)=n$ \cite{Geiges,Etnyre}. Such submanifolds are known as Legendrian and they can be characterized through the embedding $\psi:L\,\hookrightarrow\,\mathcal{B}$, whose differential $\psi_{*}:TL\rightarrow \Pi$ is an injective application between tangent spaces. This indicates the integral nature of $L$ and it simply refers to the fact that every smooth curve over $L$ with non-zero velocity will be mapped into a smooth curve tangent to $\Pi$, also with non-zero velocity. Another way to see this condition is to employ the description of distributions in terms of contact 1-forms that lead to the expression
\begin{equation}
\label{eq:12}
\psi^{*}\,\alpha\,=\,0.
\end{equation}

The reason why we can identify $\mathcal{E}$ as a Legendrian submanifold of the TPS is because it is always possible to find coordinates $(q^{1},...,q^{n},p_{1},...,p_{n})$ called Darboux coordinates \cite{Geiges,Etnyre}, such that the contact 1-form can be expressed as  $\alpha=d\phi-\sum_{i=1}^{n}p_{i}dq^{i}$ and the embedding generates thermodynamic relations, i.e., they assign $\phi=\phi(q^{1},..,q^{n})$ the character of a thermodynamic potential
\begin{equation}
\label{eq:13}
\begin{array}{c}
\psi^{*}\,\alpha\,=\,d\left(\phi(q^{1},...,q^{n})\right)\,-\,\displaystyle\sum\limits_{i=1}^{n}p_{i}(q^{1},...,q^{n})dq^{i}\,=\,0\\
\\
\Downarrow\\
\\
\psi:(q)\mapsto\left(q,p=\displaystyle\frac{\partial \phi}{\partial q},\phi(q)\right).
\end{array}
\end{equation}
We can think of $\mathcal{B}$ as a $2n+1$-dimensional manifold where each embedding of a smooth $n$-dimensional manifold describes a thermodynamic system. 

Darboux coordinates are not unique. Indeed, there exists an infinite number of coordinate changes that map the functional form $\alpha=d\phi-\sum_{i=1}^{n}p_{i}dq^{i}$ in coordinates $(q^{1},...,q^{n},p_{1},...,p_{n},\phi)$ into the same expression with new coordinates $(q^{'1},...,q^{'n},p'_{1},...,p'_{n},\phi')$ \cite{Geiges,Etnyre}. A particular method to generate Darboux coordinates consists in applying Legendre transformations (\ref{eq:4}), which can be represented as a discrete transformation of coordinates on $\mathcal{B}$ \cite{Arnold}
\begin{equation}
\label{eq:14}
\mathcal{L}_{I}(q,p,\phi)=\left\{\begin{array}{l}
q^{'I}=p_{I},\\
q^{'J}=q^{J},\\
p'_{I}=-q^{I},\\
p'_{J}=p_{J},\\
\phi'=\phi-\displaystyle\sum\limits_{k\in I}q^{k}p_{k}\end{array}\right.
\end{equation}

Legendre transformations can be interpreted as changes of coordinates on $\mathcal{B}$ that preserve the contact distribution $\Pi$. Diffeomorphisms on $\mathcal{B}$ that satisfy this property are known as contactomorphisms that turn out to form a Lie group  \cite{Geiges,Etnyre}, called contact group $\textup{Cont}(\mathcal{B},\Pi)\equiv \left(\left\{F\in \textup{Diff}(\mathcal{B}):\,F_{*}\Pi=\Pi\right\},\circ \right)$. In this way, every contactomorphism relates Darboux coordinate systems. 

Another important feature of the TES has to do with quasi-static thermodynamic processes, which are continuous successions of equilibrium states \cite{Callen}, i.e.,   smooth curves over the embedding of $\mathcal{E}$ in $\mathcal{B}$ \cite{Quevedo,Quevedo2}. Particularly, since $\mathcal{E}$ is a Legendrian submanifold, its curves are always tangent to the distribution $\Pi$. In contact geometry, these curves are known as horizontal curves \cite{Geiges,Etnyre}.

Considering that two thermodynamic equilibrium states are always connected through a quasi-static process (without considering the sense of orientation of that process) \cite{Callen}, it results necessary that any two points in a neighborhood of $\mathcal{B}$ are always connected through a horizontal curve. This is true by virtue of the Chow theorem, establishing that every bracket-generating distribution in $\mathbb{R}^{n}$  is horizontal connected \cite{Zorich}. A contact distribution $\Pi$ locally satisfies the conditions of the Chow theorem. 
This tells us that the tangential space of $\mathcal{B}$ can be constructed joining all the vector fields in $\Pi$ and all Lie brackets between them\footnote{$\Pi$ is said to be bracket-generating of step 2. In general a bracket-generating distribution $\mathcal{D}$ of step $n$ is one that satisfies $T\mathcal{B}=[\mathcal{D},\mathcal{D}]^{n}$, where $[\mathcal{D},\mathcal{D}]^{1}=\mathcal{D}$, $[\mathcal{D},\mathcal{D}]^{2}=\mathcal{D}^{1}+[\mathcal{D}^{1},\mathcal{D}]$,..., $[\mathcal{D},\mathcal{D}]^{n}=\mathcal{D}^{n-1}+[\mathcal{D}^{n-1},\mathcal{D}]$ \cite{Zorich,Arkady}.}:
\begin{equation}
\label{eq:15}
T\mathcal{B}\,=\,\Pi\oplus[\Pi,\Pi],
\end{equation}
\noindent where $[\Pi,\Pi]$ is the vector space generated by the Lie bracket between all the elements of $\Pi$ \cite{Zorich,Arkady}. Indeed, to prove (\ref{eq:15}), we have to notice that from the vector field induced by Darboux coordinates we can build the basis $(\frac{\partial}{\partial p_{1}},...,\frac{\partial}{\partial p_{n}},\frac{\partial}{q^{1}}+p_{1}\frac{\partial}{\partial \phi},...,\frac{\partial}{q^{n}}+p_{n}\frac{\partial}{\partial \phi})$, whose bracket generates the only complementary vector field to $\Pi$ on $T\mathcal{B}$. In this case, $\frac{\partial}{\partial \phi}$: $\,\left[\frac{\partial}{\partial p_{i}},\frac{\partial}{\partial q^{i}}+p_{i}\frac{\partial}{\partial \phi}\right]\,=\,\frac{\partial}{\partial \phi}\,\Longrightarrow\, \left[\Pi,\Pi\right]\,=\,\textup{span}\left(\frac{\partial}{\partial \phi}\right)$.

\section{Frame-bundle structure of the thermodynamic phase space}
\label{sec:bundle}

As shown in \cite{Mrugala1991}, the thermodynamic phase space $\mathcal{B}$ has a principal fiber bundle structure where the 1-form contact gives rise to a principal connection, i.e., a 1-form over $\mathcal{B}$ with values in the Lie algebra of the structure Lie group.

To better develop this idea, it is important to understand that in a contact manifold the existence of a vector field $\xi$ is guaranteed with the characteristic that, for every point $z\in \mathcal{B}$, $\xi(z)$ is not within the distribution at that point, i. e., $\xi(z)\not\in \Pi(z)$. When we choose $\alpha$ as a representative of the 1-form contact family, this vector field $\xi$ can be defined exactly from the conditions
\begin{eqnarray}
\label{eq:condicion_1}
i_{\xi}\alpha&=&1,\\
\label{eq:condicion_2}
i_{\xi}d\alpha&=&0,
\end{eqnarray}
\noindent where (\ref{eq:condicion_2}) arises as a direct consequence of demanding for (\ref{eq:condicion_1}) the condition of maximal non-integrability (\ref{eq:11}). Locally, using the tangent base $(\frac{\partial}{\partial \phi},\frac{\partial}{\partial q^{1}},...,\frac{\partial}{ \partial q^{n}},\frac{\partial}{\partial p_{1}},...,\frac{\partial}{\partial p_{n}})$ generated from a set of Darboux coordinates, we can make the identification $\xi=\frac{\partial}{\partial \phi}$.

Given the one-dimensional foliation generated by the integrable curves of the $\xi$ field, called the Reeb field, we find that $\mathcal{B}$ is the total space of a fiber bundle over the quotient space $M\equiv\mathcal{B}/\xi$, where the fibers are isomorphic to the additive group of real numbers, i. e., $G=(\mathbb{R},+)$ \cite{Mrugala1991}. In this scheme, it is easy to see that the 1-form $\alpha\otimes \xi$ becomes a principal connection on $\mathcal{B}$, that is, a 1-form over $\mathcal{B}$ with values in the Lie algebra $\mathfrak{g}=\textup{span}(\xi)$. Indeed, under the right-action of $G$ on the fibers, i. e., $R_{a}(z)=z\triangleleft a=(\phi+a,q,p)$, in Darboux coordinates, $\alpha$ transforms appropriately with the Adjoint action of the group $G$, taking into account that the said action is trivial since $G$ is an abelian group:
\[\begin{array}{ll}
R_{a}^{*}(\alpha\otimes \xi)&=R_{a}^{*}(d\phi-\displaystyle\sum\limits_{i=1}^{n}p_{i}dq^{i})\otimes \xi\\
&=(d(\phi+a)-\displaystyle\sum\limits_{i=1}^{n}p_{i}dq^{i})\otimes \xi\\
&=(d\phi-\displaystyle\sum\limits_{i=1}^{n}p_{i}dq^{i})\otimes \xi=\alpha\otimes \xi\\
&=\alpha\otimes \textup{Ad}_{a}\xi=\textup{Ad}_{a}(\alpha\otimes \xi).
\end{array}\]
\noindent Furthermore, also trivially identifying the fundamental fields $(aR)^{\#}=\frac{d}{dt}R_{a t}=a R$, we have the evaluation
\[i_{(aR)^{\#}}(\alpha\otimes R)=(i_{aR}\alpha)R=(ai_{R}\alpha)R=aR.\]

An important consequence of the fact that the contact 1-form can be seen as a principal connection is that the horizontal distribution generated by $\alpha\otimes \xi$ from its kernel coincides precisely with the contact distribution of the kernel of $\alpha$, in such a way that the horizontal lifts of curves on $M$  are always tangent to $\Pi$. Using the Darboux coordinates $(\phi,q,p)$ in $\mathcal{B}$, the horizontal lift of a curve $\gamma:s\in [0,1]\rightarrow (q(s) ,p(s))\in M$ has the form $\widetilde{\gamma}:s\in [0,1]\rightarrow (\phi(s),q(s),p(x))\in \mathcal{B}$ for $\phi:[0,1]\rightarrow \mathbb{R}$, solution of the horizontal lifting equation $i_{\frac{d\widetilde{\gamma} }{ds}}\alpha=0$, i. e.,
\begin{equation}
\label{eq:lev_hor}
\phi(s)=\displaystyle\int_{0}^{s}\!\!d\tau\,p_{i}(\tau)\frac{dq^{i}}{d\tau}\ .
\end{equation}

\noindent Now, taking into account that any vector field tangent to a Legendrian submanifold is, by definition, a contact field, a Legendrian submanifold is completely foliated by horizontal curves and can thus be understood as a horizontal lifting of a $n$-dimensional submanifold in $M$. We know precisely that in these Legendrian curves, equations-type of state $p_{i}(q)=\frac{\partial \phi}{\partial q^{i}}$ are satisfied, that when substituted in (\ref {eq:lev_hor}) give rise to the tautology
\[\begin{array}{c}
\phi(s)=\displaystyle\int_{0}^{s}d\tau \displaystyle\frac{\partial \phi}{\partial q^{i}}\displaystyle\frac{d q^{i}}{d\tau}=\displaystyle\int_{\gamma}d\phi=\phi\circ \gamma(s)\\
\\
\Downarrow\\
\\
\widetilde{\gamma}=(\phi(s),q(s),\frac{\partial \phi}{\partial q}(s)).\end{array}\]
\noindent In particular, for a closed loop $\gamma=\partial C$ in $M$, which encloses a region $C\subset M$, note that the holonomy of its horizontal lift, i. e., the vertical difference (on the fiber) between the initial and final value of the lifted curve, is given by the integral
\[\Delta \phi=\int_{0}^{1}ds\,p_{i}(s)\frac{dq^{i}}{ds}=\int_{\partial C}\theta=\int_{C}d\theta ,\]
\noindent where $\theta=\sum_{i=1}^{n}p_{i}dq^{i}$ is a 1-form over the open subset of $M$ containing $\gamma$ with  coordinates $(q,p)$. Identifying, at least locally, $d\theta=\sum_{i=1}^{n}dp_{i}\wedge dq^{i}$ as a symplectic 2-form in $C$, we see that the condition that 
horizontal lifted curves close, as it is necessary for Legendrian curves, suggests the existence of a symplectic structure in $M$.

\section{Contactization of the cotangent bundle}
\label{sec:contact}

The introduction of the TPS $\mathcal{B}$  as a geometric environment of the TES  $\mathcal{E}$ is not enough to fix the topology of $\mathcal{B}$, although requesting that $\mathcal{B}$ be a contact manifold is a quite restrictive condition. This allows us to consider different candidates for the topology of $\mathcal{B}$. We will now present a specific candidate that is particularly interesting because it is naturally built from the TES.

Let be $T^{*}\mathcal{E}$ the cotangent bundle of the manifold $\mathcal{E}$, defined as the vector fiber bundle with base space $\mathcal{E}$, fibers given as the   cotangent spaces, and the natural projection $\pi:T^{*}\mathcal{E}\rightarrow \mathcal{E}$ \cite{Nakahara}. In terms of the cotangent bundle, the 1-forms on $\mathcal{E}$, $\omega \in \Lambda^{1}(\mathcal{E})$, are interpreted as embeddings of the mentioned TES in $T^{*}\mathcal{E}$ \cite{Nakahara}: $\omega:\mathcal{E}\,\rightarrow\,T^{*}\mathcal{E}$ with $\pi\circ \omega=\textup{id}_{\mathcal{E}}$. In fact, if $\omega \in \Lambda^{1}(\mathcal{E})$ has components $(p_{1},...,p_{n})$ in the cotangent basis $(dq^{1},...,dq^{n})$, i.e. $\omega=\sum_{i=1}^{n}p_{i}dq^{i}$, the  $2n$ real numbers $(q^{1},...,q^{n},p_{1},...,p_{n})$ turn out to be a natural coordinate system for $T^{*}\mathcal{E}$.

An important feature of any cotangent bundle is that it has a canonical 1-form $\theta\in \Lambda^{1}(T^{*}\mathcal{E})$ \cite{Arnold,Silva}, defined at any point $\forall\,(x,\omega) \in T^{*}\mathcal{E}$ as
\begin{equation}
\label{eq:16}
\begin{array}{c|} 
\theta
\end{array}_{\,\,(x,\omega)}\,\equiv\,\begin{array}{c|}\left(\pi^{*}\omega\right)\end{array}_{\,\,x},
\end{equation}
\noindent whose exterior derivative defines a symplectic form on $T^{*}\mathcal{E}$: $\Omega\equiv d\theta$ \cite{Arnold,Silva}. Accordingly, we say that every cotangent bundle has an exact symplectic structure in it. Another remarkable property of the canonical 1-form $\theta$ is the so-called reproduction property for any $\omega\,\in\,\Lambda^{1}(\mathcal{E})$:
\begin{equation}
\label{eq:17}
\omega^{*}\theta\,=\,\omega,
\end{equation}
\noindent 
indeed, $\omega^{*}\theta=\omega^{*}\left(\pi^{*}\omega\right)=\,\left(\pi\circ \omega\right)^{*}\omega\,=\,\omega$.
We note that in the coordinate system $(q^{1},...,q^{n},p_{1},...,p_{n})$ that we have just described, the 1-form $\theta$ and the 2-form $\Omega$ acquire their simplest expression $\theta=\sum_{i=1}^{n}p_{i}dq^{i}$ and $\Omega=\sum_{i=1}^{n}dp_{i}\wedge dq^{i}$.

Considering the close relationship between a contact structure and a symplectic structure, we can associate $T^{*}\mathcal{E}$ with a special construction that we will call its contactization $\mathcal{B}_{C}$, and which is roughly speaking a (not-unique)\footnote{Although this construction is not unique, for the purposes of this article this does not prove to be a problem. This question is treated in the Appendix.} lineal and trivial fiber bundle over $T^{*}\mathcal{E}$ \cite{Geiges,Etnyre,Feng} with natural projection given by $\pi_{C}:\,\mathcal{B}_{C}\equiv T^{*}\mathcal{E}\times \mathbb{R}\,\rightarrow\,T^{*}\mathcal{E}$. A special feature of the  contactization $\mathcal{B}_{C}$ is that it contains a natural, co-orientable and maximally non-integrable contact structure, built from the symplectic structure of $T^{*}\mathcal{E}$, described by the family of contact 1-forms $[\alpha_{C}]$ where $\alpha_{C}\,=\,dz-\pi_{C}^{*}\theta$ with $z\in \mathbb{R}$ as the fiber coordinate. Indeed, let us take the top-form over $\mathcal{B}_{C}$ given by $\alpha_{C}\wedge d\alpha_{C}^{\wedge n}=-dz\wedge \pi_{C}^{*}(\Omega^{\wedge n})$, and let  $(\frac{\partial}{\partial z},X_{1},...,X_{n})$  be a vector field basis for $T\mathcal{B}_{C}$. If we evaluate this basis in $\alpha_{C}\wedge d\alpha_{C}^{\wedge n}$, we obtain $-\Omega^{\wedge n}(\pi_{C*}X_{1},...,\pi_{C*}X_{2n})\neq 0$, because $\Omega^{\wedge n}$ is not null at all $T^{*}\mathcal{E}$ and represents the  so-called Liouville volume top-form \cite{Arnold,Silva}. Thus, we see that $\alpha_{C}$ satisfies condition (\ref{eq:12}) and is, therefore,  a contact 1-form.
 
An interesting fact about the contactization of an exact symplectic manifold is that it builds a bridge between symplectic geometry and contact geometry. We can highlight that all contactomorphisms $F$ on $\mathcal{B}_{C}$ result, on any embedding $\ell :T^{*}\mathcal{E}\hookrightarrow \mathcal{B}_{C}$ with $\pi_{C}\circ \ell=\textup{id}_{T^{*}\mathcal{E}}$, in a diffeomorphism on $T^{*}\mathcal{E}$ that leaves invariant the symplectic form $\Omega$, i.e., 

\[\begin{array}{l}
\left(\pi_{C}\circ F\circ \ell\right)^{*}\Omega=\ell^{*}\circ F^{*}\circ \pi_{C}^{*}d\theta\\
\\
\phantom{\left(\pi_{C}\circ F\circ \ell\right)^{*}\Omega}=-\ell^{*}\circ F^{*}(d\alpha_{C})\\
\\
\phantom{\left(\pi_{C}\circ F\circ \ell\right)^{*}\Omega}=-\ell^{*}d\left(F^{*}\alpha_{C}\right)=-\ell^{*}d\alpha_{C}\\
\\
\phantom{\left(\pi_{C}\circ F\circ \ell\right)^{*}\Omega}=-\ell^{*}\circ \pi_{C}^{*}\Omega=\left(\pi_{C}\circ \ell\right)^{*}\Omega=\Omega.
\end{array}\]

As long as we can embed $\mathcal{E}$ into $T^{*}\mathcal{E}$, such that $\Omega$ is zero on its tangent spaces, we can construct a Legendrian embedding of $\mathcal{E}$ into $\mathcal{B}_{C}$ using any cross section of this last fiber bundle. First of all, note that for any 1-form $\gamma\in \Lambda^{1}(\mathcal{E})$ and $\ell:T^{*}\mathcal{E}\rightarrow \mathcal{B}_{c}$ with $\pi_{C}\circ \ell=\textup{id}_{T^{*}\mathcal{E}}$, the composition $\psi_{\gamma}\equiv \ell\circ \gamma$ is an embedding of $\mathcal{E}$ into $\mathcal{B}_{C}$. Note that if $\psi_{\gamma}$ is a Legendrian map, as defined in Eq.(\ref{eq:13}), then $\gamma$ turns out to be an exact 1-form:

\[\begin{array}{l}
\psi_{\gamma}^{*}\alpha_{C}\!=\!d(\psi_{\gamma}^{*}z)\!-\!\psi_{\gamma}^{*}\circ \pi_{C}^{*}\theta\\
\\
\phantom{\psi_{\gamma}^{*}\alpha_{C}}\!=\!d(\psi_{\gamma}^{*}z)\!-\!(\pi_{C}\circ \ell\circ \gamma)^{*}\theta\\
\\
\phantom{\psi_{\gamma}^{*}\alpha_{C}}\!=\!d(\psi_{\gamma}^{*}z)\!-\!\gamma^{*}\theta\!=\!d(\!\psi_{\gamma}^{*}z\!)-\!\gamma\!=\!0\Longrightarrow \gamma\!=\!d(\!\psi_{\gamma}^{*}z\!),
\end{array}\]

\noindent and therefore $\gamma$ embed $\mathcal{E}$ into $T^{*}\mathcal{E}$ as a Lagrangian  submanifold, that is, an $n$-dimensional submanifold where the constraint of the symplectic form $\Omega$ is zero:

\begin{equation}
\label{eq:18}
\gamma^{*}\Omega\,=\,\gamma^{*}\left(d\theta\right)\,=\,d\left(\gamma^{*}\theta\right)\,=\,d\gamma\,=\,d^{2}\left(\psi_{\gamma}^{*}z\right)\,=\,0.
\end{equation}

Lagrangian submanifolds are interesting and well-studied elements of symplectic geometry \cite{Arnold,Silva,Weinstein}. Among their properties, we can highlight that given any smooth function $h\in \mathcal{C}^{\infty}(T^{*}\mathcal{E})$, which we call a Hamiltonian, the flow generated by its associated Hamiltonian field, that is, the vector field $X_{h}:T^{*}\mathcal{E}\rightarrow T\left(T^{*}\mathcal{E}\right)$, identified with $dh$ by means of $\Omega$ \cite{Arnold,Silva},

\begin{equation}
\label{eq:19}
dh\,=\,i_{X_{H}}\Omega,
\end{equation}
maps a Lagrangian submanifold to another Lagrangian submanifold. Indeed, given the definition (\ref{eq:19}) it is easy to show that the flow of $X_{h}$ leaves both $h$ and $\Omega$ invariant\footnote{$\mathscr{L}_{Z}$ is the Lie derivative with respect the flow of the vector field $Z$.}: 

\begin{equation}
\label{eq:20}
\begin{array}{l}
\mathscr{L}_{X_{h}}h = \langle dh,X_{h}\rangle=\Omega(X_{h},X_{h})=0.\\
\\
\mathscr{L}_{X_{h}}\Omega=\left(i_{X_{h}}d+di_{X_{h}}\right)\Omega=d\left(i_{X_{h}}\Omega\right)=d^{2}h=0.
\end{array}
\end{equation}

Particularly, if a Hamiltonian $h$ is constant along a Lagrangian submanifold $\mathcal{E}$, $\gamma^{*}h=\textup{const}$, then the flow of the Hamiltonian vector field maps the Lagrangian submanifold onto itself \cite{Vallina}. This can be seen from the fact that $X_{h}$ is always tangent to $\mathcal{E}$ since $X_{h}$ is symplectically orthogonal to every vector field $Y$ tangent to $\mathcal{E}$, that is,
\begin{equation}
\label{eq:21}
\Omega(X_{h},\gamma_{*}Y)\!=\!i_{Y}\left(\gamma^{*}dh\right)\!=\!i_{Y}d\left(\gamma^{*}h\right)\!=\!i_{Y}(0)=0,
\end{equation}
\noindent and in a Lagrangian submanifold one finds, by definition, the largest possible number of symplectically orthogonal tangent fields between each other\footnote{A Lagrangian submanifold is a maximally isotropic submanifold of a symplectic manifold \cite{Arnold,Silva,Weinstein}.}.


\subsection{Hamilton-Jacobi-like equation for thermodynamic systems}

If we take as a TPS candidate the contactization of the cotangent bundle of the TES, as we mentioned earlier\footnote{It is enough to ask that they be contactomorphic to each other, in other words, that there is a contactomorphism from the contact structure of the TPS to the contact structure of $T^{*}\mathcal{E}\times \mathbb{R}$. Locally such contactomorphism always exists \cite{Geiges,Etnyre} and the discussion reduces  to choosing a particular Darboux coordinate system.}, we can identify thermodynamic systems first as a Legendrian embedding of $\mathcal{E}$ into $\mathcal{B}\simeq \mathcal{B}_{C}$, and then as a Lagrangian embedding of $\mathcal{E}$ into $T^{*}\mathcal{E}$. This perspective allows us to see all the mappings of the cotangent bundle to $\mathbb{R}$, which are constant on $\mathcal{E}$, as generators of thermodynamic processes through the flows generated by their associated Hamiltonian vector fields. This is in  direct analogy to the use of functions and Hamiltonian vector fields of contact, defined on the TPS,  that are used in the framework of contact geometry \cite{Mrugala2, Mrugala3}.

If we start identifying the Legendrian map $\psi_{\gamma}=\ell\circ \gamma$, we note that $\gamma$ must be the exterior derivative of the thermodynamic potential, $\gamma=d(\psi_{\gamma}^{*}z)=d\phi$, which in turns leads us to propose the equation
\begin{equation}
\label{eq:22}
\phi=\psi_{\gamma}^{*}z+\phi_{0}=K(d\phi)-h_{0},
\end{equation}
\noindent where $h_{0}$ is a real constant and  we define $K$ as the portion of the lifting performed by $\ell$ of $T^{*}\mathcal{E}$ to $\mathcal{B}$, which falls on the fibers: $K\equiv z\circ \ell$. In  coordinates $(q^{1},...,q^{n})$ on $\mathcal{E}$, Eq.(\ref{eq:22}) takes the form of a Hamilton-Jacobi type equation independent of the evolution parameter \cite{Arnold,Cahill},
\begin{equation}
\label{eq:23}
\phi(q^{1},...,q^{n})\,=\,K\left(q^{1},...,q^{n},\displaystyle\frac{\partial \phi}{\partial q^{1}},...,\displaystyle\frac{\partial \phi}{\partial q^{n}}\right)-h_{0}.
\end{equation}

The similarity to the Hamilton-Jacobi equation becomes much more evident when we define the Hamiltonian function $h\equiv K-\pi^{*}\phi$, constant on $\mathcal{E}$,
\begin{equation}
\label{eq:24}
h\left(q^{1},...,q^{n},\displaystyle\frac{\partial \phi}{\partial q^{1}},...,\displaystyle\frac{\partial \phi}{\partial q^{n}}\right)=h_{0}.
\end{equation}

Thus, we find a Hamiltonian description of  thermodynamic processes, characterized by all the maps $T^{*}\mathcal{E}\rightarrow \mathbb{R}$, which are constants over the embedding equilibrium space $\mathcal{E}$.

\subsection{Hamiltonian thermodynamics of ideal gases}

To show the compatibility of symplectic thermodynamics with the already well-established contact thermodynamics, let us consider thermodynamic processes similar to those presented in \cite{Mrugala3} for ideal gases, but by means of Hamiltonian vector fields. Then, consider again the physical coordinate system $(U,V)$ on $\mathcal{E}$ and the corresponding local coordinates $(U,V,p_{U},p_{V})$ in $T^{*}\mathcal{E}$. In particular, let us work with an ideal gas with a fixed number $N$ of particles and a fundamental equation given by Eq. (\ref{eq:9}). This means that instead of working on all $\mathcal{B}_{C}$, we limit ourselves only to the region described by the embedding
\begin{equation}
\label{eq:25}
dS(U,V)=\left(U,V,p_{U}=\displaystyle\frac{N k_{B} C_{V}}{U},p_{V}=\displaystyle\frac{N k_{B}}{V}\right),
\end{equation}
\noindent where $p_{U}\equiv\frac{1}{T}$ and $p_{V}\equiv\frac{P}{T}$.

In this region, note that the maps  $h_{1}\equiv (p_{V}V)^{\alpha}$ and $h_{2}\equiv (p_{U}U)^{\alpha}$, $\alpha \geq 1$, are constants:
\begin{equation}
\label{eq:26}
\begin{array}{l}
(dS)^{*}h_{1}=\left(\displaystyle\frac{PV}{T}\right)^{\alpha}=(Nk_{B})^{\alpha}=\textup{const}.\\
\\
(dS)^{*}h_{2}=\left(\displaystyle\frac{U}{T}\right)=(Nk_{B}C_{V})^{\alpha}=\textup{const}.,
\end{array}
\end{equation}
\noindent which is the reason why the integral curves $\gamma_{1}:\mathbb{R}\rightarrow dS(\mathcal{E})$ and $\gamma_{2}:\mathbb{R}\rightarrow dS(\mathcal{E})$, associated with its Hamiltonian vector fields $X_{h_{1}}\circ \gamma_{1}(t)=\dot{\gamma}_{1}$ and $X_{h_{2}}\circ \gamma_{2}(t)=\dot{\gamma}_{2}$, respectively, describe an expansion with constant energy and temperature, and an isocoric process
\begin{equation}
\label{eq:27}
\begin{array}{l}
\gamma_{1}(t)=\left(U,Ve^{\beta t},\displaystyle\frac{1}{T},\displaystyle\frac{P}{T}e^{-\beta t}\right),\\
\\ 
\gamma_{2}(t)=\left(Ue^{\beta C_{V}^{\alpha-1}t},V,\displaystyle\frac{1}{T}e^{-\beta C_{V}^{\alpha-1}t},\displaystyle\frac{P}{T}\right),
\end{array}
\end{equation}
\noindent where we define the parameter $\beta\equiv \alpha(N k_{B})^{\alpha-1}$. First of all, we notice that effectively both $\gamma_{1}(t)$ and $\gamma_{2}(t)$ describe quasi-static processes; therefore, at each value of the parameter $t \in \mathbb{R}$, the equations of state are satisfied, i.e.,  the curves do not leave the embedding of $\mathcal{E}$. Furthermore, we note that along $\gamma_{1}$ the entropy of the ideal gas always increases as expected for a process at constant energy. In both cases the relation between the parameters $t$ and $S$ is an affine relation, i.e.,  $S\circ \gamma_{1}(t)=S_{0}+\beta t$ and $S\circ \gamma_{2}(t)=S_{0}+\beta C_{V}^{\alpha-1}t$, which allows us to rewrite Eq.(\ref{eq:27}) as
\begin{equation}
\label{eq:28}
\begin{array}{l}
\gamma_{1}(S)=\left(U,Ve^{(S-S_{0})},\displaystyle\frac{1}{T},\displaystyle\frac{P}{T}e^{-(S-S_{0})}\right),\\
\\ 
\gamma_{2}(S)=\left(Ue^{(S-S_{0})},V,\displaystyle\frac{1}{T}e^{-(S-S_{0})},\displaystyle\frac{P}{T}\right).
\end{array}
\end{equation}

Another constant Hamiltonian in the embedding is the map $h_ {3}\, =\,\displaystyle\frac {C_ {V}\,p_ {V} V} {U} -p_ {U} $,
\begin{equation}
\label{eq:29}
(dS)^{*}h_{3}\,=\,\displaystyle\frac{C_{V}\,P V}{T U}-\displaystyle\frac{1}{T}=0,
\end{equation}
with Hamiltonian integral curves $\gamma_ {3}$, given by the expression 
\begin{equation}
\label{eq:30}
\gamma_{3}(t)\!\!=\!\!\left(\!U\!-\!t,\!\displaystyle\frac{V}{(1-\frac{t}{U})^{C_{V}}},\!\displaystyle\frac{1}{T}\!\left(\displaystyle\frac{3U_{0}-t}{2U_{0}-t}\right)\!,\! \displaystyle\frac{P}{T}\!\left(\!1-\displaystyle\frac{t}{U_{0}}\!\right)^{C_{V}}\!\right).
\end{equation}

In this case, we are in the presence of an isentropic expansion, where the parametrization is again consistent with the physical idea that, at a constant entropy $S\circ\gamma_{3}(t)\,=\,S(t)$, the energy must decrease: $U-t$.

\subsection{Hamiltonian flow from an ideal gas to a van der Walls gas}

As we have shown, the Hamiltonian flow of any 0-form over $T^{*}\mathcal{E}$ maps Lagrangian submanifolds into Lagrangian submanifolds, which in the thermodynamic context means that Hamiltonian flows take equilibrium states of a system and relate them to the equilibrium states of another system. Both systems should have the same number of thermodynamic degrees of freedom. As a particular interesting example, consider the Hamiltonian
\begin{equation}
\label{eq:31}
h\,=\,-\displaystyle\frac{aV_{0}}{V^{2}}p_{U}+bp_{V}+\displaystyle\frac{a}{T_{0}V}.
\end{equation}
The corresponding flow  takes states of an ideal gas $(U_{0},V_{0},\frac{1}{T_{0}},\frac{P_{0}}{T_{0}})$ and for each $t\in\mathbb{R}$ relates them to states of a van der Waals gas characterized by the coordinates $\left(U(t),V(t),(\frac{1}{T})(t),(\frac{P}{T})(t)\right)$ ($0<a\ll 1,0<b\ll 1$).
Indeed, the integration of the Hamiltonian equations associated with (\ref{eq:31}) yields
\begin{equation}
\label{eq:32}
\begin{array}{l}
U(t)=U_{0}-\displaystyle\frac{at}{V_{0}+bt},\\
\\
V(t)=V_{0}+bt,\\
\\
\left(\displaystyle\frac{1}{T}\right)\!(t)=\displaystyle\frac{1}{T_{0}},\\
\\
\left(\displaystyle\frac{P}{T}\right)\!(t)=\displaystyle\frac{P_{0}}{T_{0}}-\displaystyle\frac{at}{T_{0}(V_{0}+bt)^{2}}.
\end{array}
\end{equation}

Notice that from the fundamental equation and equations of state of the ideal gas, both the fundamental equation of a van der Waals gas and its equations of state \cite{Callen} can be obtained as
\begin{equation}
\label{eq:33}
\begin{array}{l}
S_{0}=Nk_{B}\ln\left(U_{0}^{C_{V}}V_{0}\right)\\
\\
\phantom{S_{0}}=Nk_{B}\ln\left(\left[U(t)+\displaystyle\frac{a t}{V(t)}\right]^{C_{V}}(V(t)-bt)\right),\\
\\
P_{0}V_{0}= \left(P(t)+\displaystyle\frac{at}{V(t)^{2}}\right)(V(t)-bt)\\
\\
\phantom{P_{0}V_{0}}= Nk_{B}T(t)=Nk_{B}T_{0}.
\end{array}
\end{equation}
{Notice that the parameter $t$ has been used here to denote different equilibrium states along a quasi-static process that connects states either of the same system or of different thermodynamic systems. 
In this sense, the parameter $t$ resembles the role of an affine parameter along a geodesic. Under certain circumstances, it can be interpreted as a ``time" parameter that connects different equilibrium states of the same quasi-static process, without breaking the fundamentals of equilibrium thermodynamics, i.e., the process occurs slowly enough for the system to remain in internal physical equilibrium.   

In the above examples, we introduced heat capacities as constants that enter the equations of state of simple thermodynamic systems. Nevertheless, they could also be considered in the context of exotic systems like black holes and cosmological models \cite{lq1,aclq1}. This means that heat capacities could also be negative or even zero, implying that the physical properties of the corresponding thermodynamic systems could change drastically. For instance, a negative heat capacity seems to lead to contradictory results in the case of simple laboratory systems; however, for systems in which long-range interactions are involved, as in the case of gravitational configurations, exotic heat capacities are perfectly well defined and do not lead to physical contradictions. In general, the heat capacities introduced in this work are limited in their values only by the kind of systems they are associated to.  }

The relationships presented above between different thermodynamic systems have also been obtained in \cite{Mrugala3} by using the contact geometry structure of the TPS. Also, they have been applied in the context of black holes in \cite{gb19}.

\subsection{Hamiltonian thermodynamics of black holes}

Since 1973, when Bekenstein discovered that the horizon area $A$ of a black hole behaves as the entropy $S$ of a classical thermodynamic system \cite{bekenstein}, the 
so-called thermodynamics of black holes has been subject of intensive research \cite{bardeen, hawking, davies}.

The fundamental equation that relates the mass $M$, the angular momentum $J$ and the electric charge $Q$ of a Kerr-Newman black hole with the entropy $S$ (proportional to the horizon area as $S=A/4$) is given by
\begin{equation}
\label{bh_entropy}
S=2\pi \left(M^{2}-\displaystyle\frac{Q^{2}}{2}+ 
R \right),
\end{equation}
\noindent where for convenience we define 
\begin{equation}
\label{R_function}
R\equiv \sqrt{M^{4}-Q^{2}M^{2}-J^{2}}.
\end{equation}

According to the first law of black hole thermodynamics \cite{bardeen}
\begin{equation}
\label{bh_first_law}
dS= \displaystyle\frac{1}{T} dM-\displaystyle\frac{\phi}{T}dQ-\displaystyle\frac{\Omega_{H}}{T}dJ,
\end{equation}

\noindent the thermodynamic dual variables are the Hawking temperature $T$, which is proportional to the surface gravity on the horizon, the electric potential $\phi$, and the angular velocity $\Omega_{H}$ on the horizon. Using (\ref{bh_entropy}) and the equations of state we find that $T$, $\phi$, and $\Omega_{H}$ are given as
\begin{eqnarray}
\label{bh_T}
T=\displaystyle\frac{R}{2 M S},\quad \phi=\displaystyle\frac{Q(S+\pi Q^{2})}{2 MS},\quad \Omega_{H}=\displaystyle\frac{\pi J}{M S}.
\end{eqnarray}

Consider the space of equilibrium states $\mathcal{E}$ of a black hole with the coordinate system $(M,Q,J)$. In particular consider the region 
\[A=\{(M,Q,J)\in \mathcal{E}: M>0, R(M,Q,J)\geq 0\},\]

\noindent and its embedding $dS(A)\subset T^{*}\mathcal{E}$ under the cross section
\begin{equation}
\label{bh_emb}
\begin{array}{ll}
dS(M,Q,J)\!=\!\!\!\!&\left(M,Q,J,\right.\\
&\left.p_{M}\!=\!\frac{2MS}{R},p_{Q}\!=\!-\frac{Q(S+\pi Q^{2})}{R},p_{J}\!=\!-\frac{2\pi J}{R}\right)\!.
\end{array}
\end{equation}

\noindent It is easily seen that over $dS(A)$ the three Hamiltonians given in (\ref{bh_hamiltonianas}) are constant; in fact, they are zero, $dS^{*}h_{1}=dS^{*}h_{2}=dS^{*}h_{3}=0$,
\begin{equation}
\label{bh_hamiltonianas}
\begin{array}{l}
h_{1}=p_{M}-\displaystyle\frac{2 M S}{R},\\
\\
h_{2}= M p_{M}-\displaystyle\frac{2 M^{2}S}{R}+2 Jp_{J}+\displaystyle\frac{4\pi J^{2}}{R},\\
\\
h_{3}=\displaystyle\frac{p_{M}}{2M}+\displaystyle\frac{p_{Q}}{2Q}+\displaystyle\frac{p_{J}}{2J}(M^{2}-Q^{2})+\displaystyle\frac{2\pi M^{2}-(S+\pi Q^{2})}{2R}.
\end{array}
\end{equation}

\noindent 
As a consequence of these Hamiltonians being zero, as we have already seen, their Hamiltonian flows are trapped on $ dS(A)$ and constitute families of thermodynamic processes represented through integral curves.

In the case of $h_{1}$, the integral curve with initial condition $\gamma_{1}(0)=dS(M,Q,J)$ is (\ref{bh_sol1}), where $\tau\in \mathbb{R}$ is a parameter with the same units of $M$, and we define  $R_{1}=R(M+\tau,Q,J)$ and $S_{1}=S(M+\tau,Q,J)$. 

\begin{equation}
\label{bh_sol1}
\begin{array}{ll}
\gamma_{1}(\tau)=\left(M+\tau,Q,J,\right.&\displaystyle\frac{2 (M+\tau)S_{1}(\tau)}{R_{1}(\tau)},\\
&\left.-\displaystyle\frac{Q(S_{1}(\tau)+\pi Q^{2})}{R_{1}(\tau)},-\displaystyle\frac{2\pi J}{R_{1}(\tau)}\right)
\end{array}
\end{equation}

\noindent We can see that this process describes a linear increase in the mass of the black hole, without any change in its electrical charge or in its total angular momentum, under which the entropy increases, but both the temperature, the electric field and the angular velocity decrease:
\[\begin{array}{l}
\displaystyle\frac{dT}{d\tau}=-\displaystyle\frac{T(\tau)}{M}\left[1+2\displaystyle\frac{M^{2}}{R(\tau)}\left(1+\displaystyle\frac{M^{2}-Q^{2}/2}{R(\tau)}\right)\right]<0,\\
\\
\displaystyle\frac{d\phi}{d\tau}=-\left[\displaystyle\frac{\phi(\tau)}{M}+\displaystyle\frac{\pi Q^{3}}{R(\tau)}\right]<0,\\
\\
\displaystyle\frac{d\Omega_{H}}{d\tau}=-\displaystyle\frac{\Omega_{H}(\tau)}{M}\left[1+\displaystyle\frac{ 2M^{2}}{R(\tau)}\right]<0.
\end{array}\]

In the case of $h_{2}$, the solutions of the Hamilton equations under the initial condition $\gamma_{2}(0)=dS(M,0,J)$ are given by 
\begin{equation}
\label{bh_sol2}
\begin{array}{ll}
\gamma_{2}(t)=\left(M e^{t},0,J e^{2t},\displaystyle\frac{2 M S}{R}e^{t},0,-\displaystyle\frac{2\pi J}{R}\right),
\end{array}
\end{equation} 
where $R_{2}=R(M e^{t},0,J e^{2t})=e^{2t}R(M,0,J)=e^{2t}R$ and $S_{2}=S(Me^{t},0,J e^{2t})=e^{2t} S(M,0,J)=e^{2t}S$, and $t\in \mathbb{R}$ is a dimensionless parameter.

\noindent In this process, as a result of an exponential increase in electrically neutral matter with non-zero angular momentum, the entropy growth is exponential and, on the contrary, both the temperature and the angular velocity decrease exponentially, whereas the electrostatic potential remains at zero, as expected,
\[\begin{array}{l}
T(t)=\displaystyle\frac{R}{2 M S}e^{-t}=Te^{-t},\\
\\
\phi(t)=0,\\
\\
\Omega_{H}(t)=\displaystyle\frac{\pi J}{M S}e^{-t}=\Omega_{H}e^{-t}.
\end{array}\]

Finally, 
\begin{equation}
\label{bh_sol3}
\begin{array}{ll}
\gamma_{3}(\lambda)\!=\!\!\!\!&(\sqrt{M^{2}+\lambda},\sqrt{Q^{2}+\lambda},\sqrt{J^{2}+(M^{2}-Q^{2})\lambda},\\
&\displaystyle\frac{2\sqrt{M^{2}+\lambda}}{R}(S+\pi \lambda),-\displaystyle\frac{\sqrt{Q^{2}+\lambda}}{R}(S\!+\!2\pi \lambda\!+\!\pi Q^{2}),\\
&\left.-\displaystyle\frac{2\pi }{R}\sqrt{J^{2}+\left(M^{2}-Q^{2}\right)\lambda},
\right).
\end{array}
\end{equation}
is the integral curve for $h_{3}$ with initial condition $\gamma_{3}(0)=dS(M,Q,J)$, where $\lambda >0$ is a parameter with the same units of $M^{2}$.

\begin{figure}
	\centering
	\includegraphics[scale=0.5]{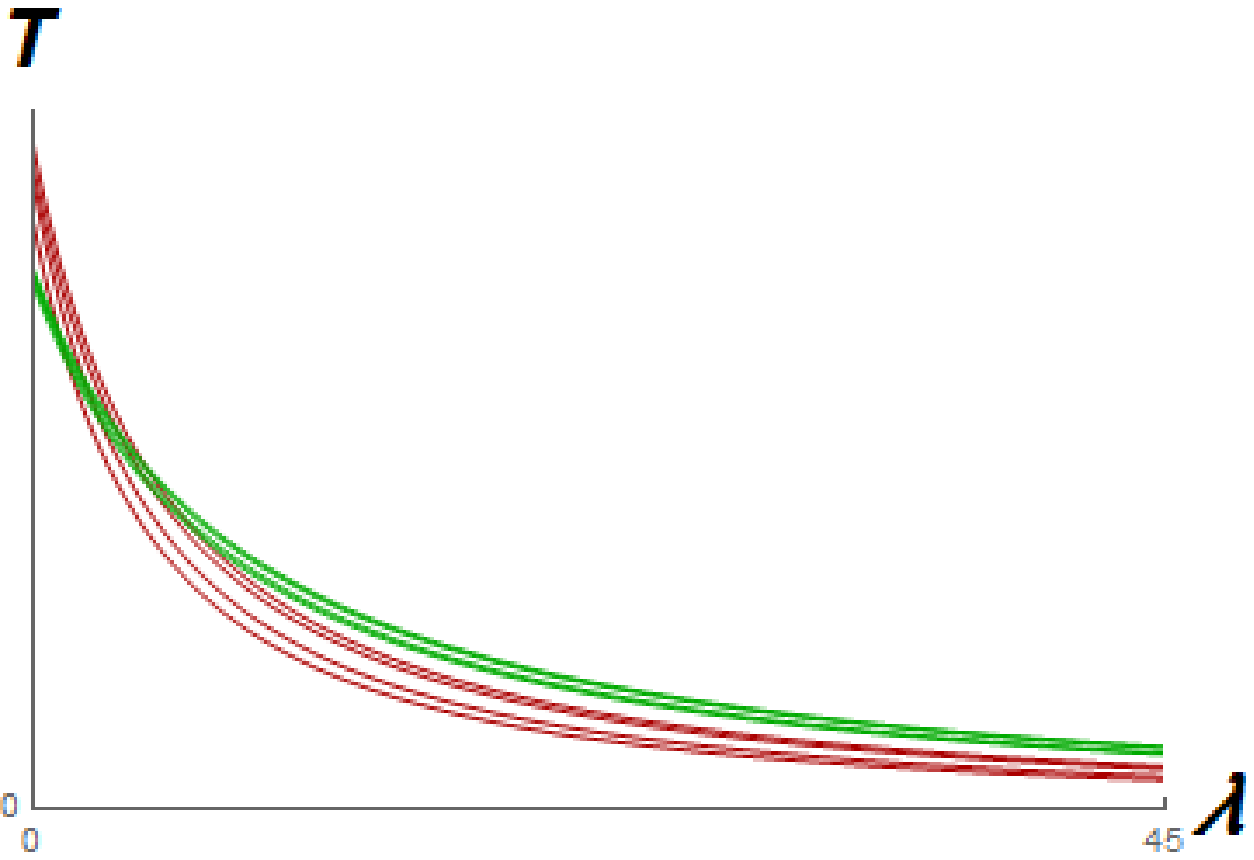}
	\includegraphics[scale=0.5]{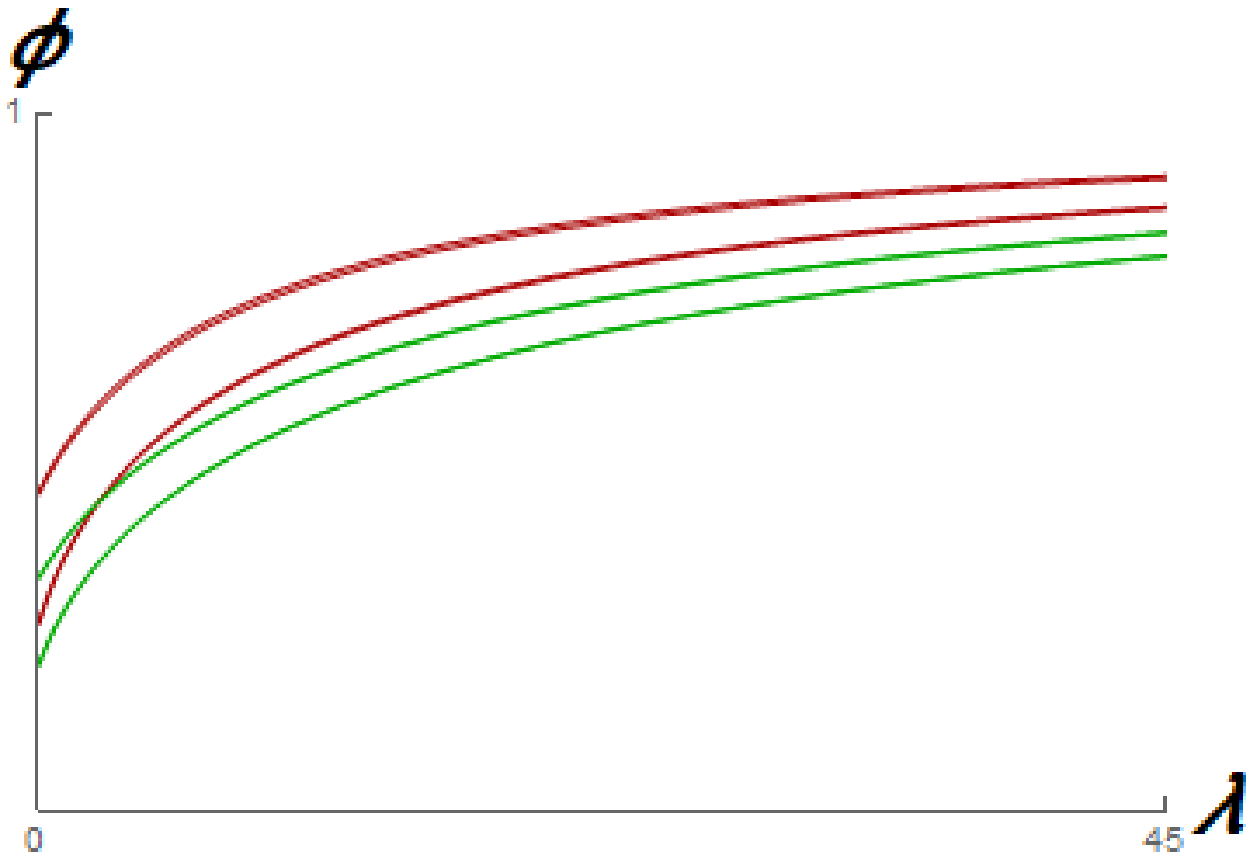}
	\includegraphics[scale=0.5]{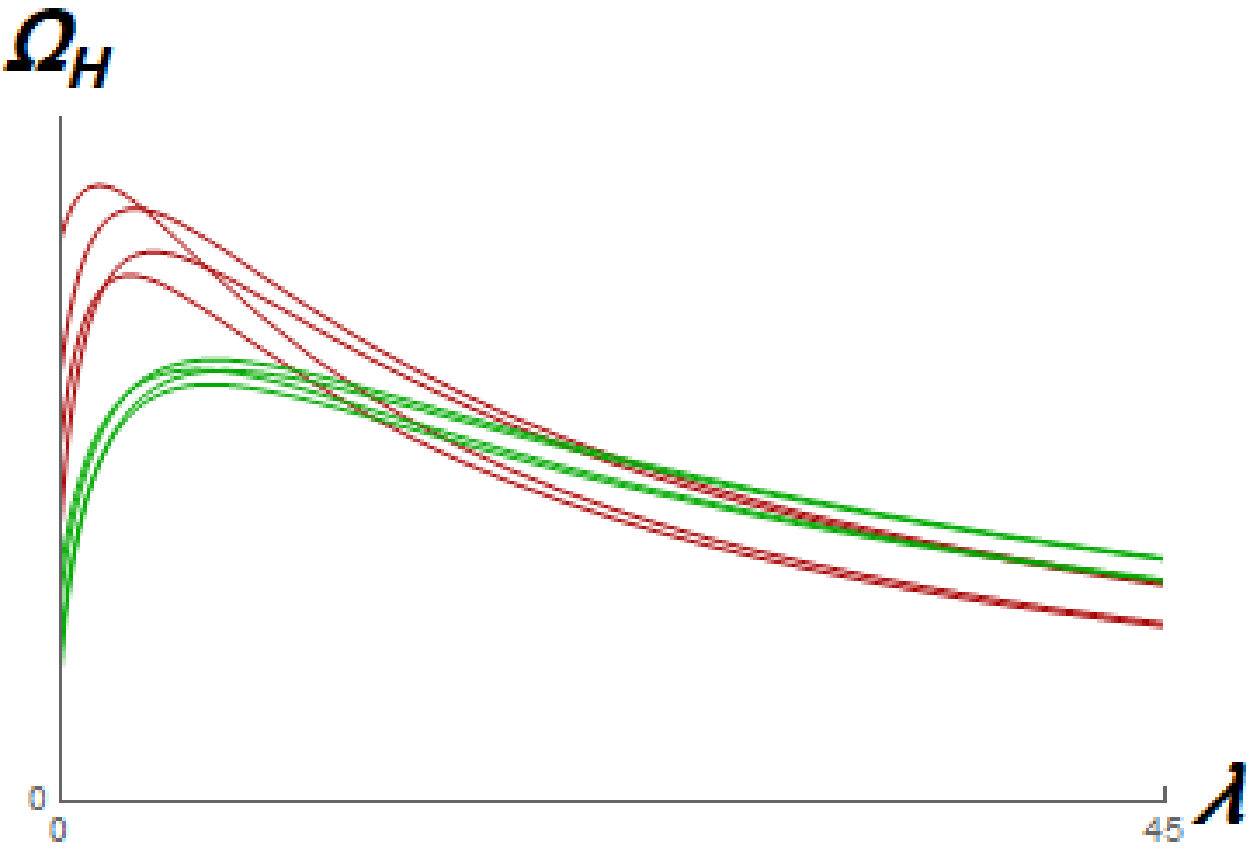}
	\caption{Behavior of $T$, $\phi$ and $\Omega_{H}$ in the $\gamma_{3}$-process, for values $(Q,M,J)$ between $(2,1,1)$ (red) and $(2.5,1.5,1.5)$ (green).}
	\label{fig1}
\end{figure}

\noindent 
The $\gamma_{3}$-related process 
\[\begin{array}{l}
\tau(\lambda)=\displaystyle\frac{R}{2 \sqrt{M^{2}+\lambda}(S+\pi \lambda)},\\
\\
\phi(\lambda)=\displaystyle\frac{1}{2}\sqrt{\displaystyle\frac{Q^{2}+\lambda}{M^{2}+\lambda}}\left(1+\displaystyle\frac{\pi(Q^{2}+\lambda)}{S+\pi \lambda}\right),\\
\\
\Omega_{H}(\lambda)=\displaystyle\frac{\pi}{S+\pi \lambda} \sqrt{\displaystyle\frac{J^{2}+(M^{2}-Q^{2})\lambda}{M^{2}+\lambda}},
\end{array}\]
describes an increase in mass, charge, and angular momentum that generates a linear increase in entropy. Indeed, $S(\lambda)=S+\pi \lambda$. 
The temperature is a monotonically decreasing function of $\lambda$, as opposed to $\phi$, which increases asymptotically until reaching the maximum value of 1 (in the appropriate units). The angular velocity experiences a brief initial growth but ends up becoming zero. This behavior is illustrated in Fig. \ref{fig1}.

\section{The Legendre group}
\label{sec:group}

Earlier we mentioned that Legendre transforms in a thermodynamic system allow us to describe thermodynamic states and processes by new control parameters, for example exchanging volume for pressure, temperature for entropy, etc. Legendre transforms do this by introducing new thermodynamic potentials by taking $\phi$ and transforming into a function $\phi'$ of the new control parameters, which contains exactly the same thermodynamic information.
In this way, in the framework of contact geometry, these transformations are implemented by means of contactmorphisms of the form (\ref{eq:14}) in the thermodynamic phase space $\mathcal{B}$. Although the subset of transformations (\ref{eq:14}) does not belong to a subgroup of contactmorphisms under composition, the goal of this section is to show that in particular one-dimensional transforms do belong to a larger subgroup of contactmorphisms, which in turn induces a group of symplectomorphisms in $T^{*}\mathcal{E}$; this after performing the identification $\mathcal{B}=\mathbb{R} \times T^{*}\mathcal{E}$. This subgroup, which we call in this work the \textit{Legendre group}, is of interest to us because it implements a symplectic action that allows us to characterize by means of $n$ constant quantities the regions whose points are different representations of the same thermodynamic state under different continuous transformations of Legendre.

Let us start by identifying $\mathcal{L}_{i}$ ($I=[i]$ for $i=1,2,..,n$) as a particular element of a one-parameter group of infinitesimal transformations: $\{\mathcal{L}_{\alpha^{i}}\}_{\alpha^{i}\in S^{1}}$, where
\[\mathcal{L}_{\alpha^{i}}(\phi,q,p)=\left\{\begin{array}{ll}
\phi'=\sin(2\alpha^{i})[\frac{(q^{i})^{2}-(p_{i})^{2}}{4}]\\
\\
\phantom{\phi'=+}-p_{i}q^{i} \sin^{2}(\alpha^{i})+\phi,\\
\\
q^{'i}=q^{i}\cos(\alpha^{i})-p_{i}\sin(\alpha^{i}),\\
\\
q^{'k}=q^{k},\\
\\
p'_{i}=q^{i}\sin(\alpha_{i})+p_{i}\cos(\alpha_{i}),\\
\\
p'_{k}=p_{k}.
\end{array}\right.\]
\noindent Specifically, $\mathcal{L}_{\alpha^{i}=\frac{\pi}{2}}=\mathcal{L}_{i}$. For each value $\alpha^{i}\in S^{1}$ these transformations $\mathcal{L}_{\alpha^{i}}$ are contact transformations \cite{Monsalvo}, which induce symplectomorphisms on $ T^{*}\mathcal{E}$ by pulling back through a cross section $\ell: T^{*}\mathcal{E}\rightarrow \mathcal{B}$ and then projecting back to $T^{*}\mathcal{E}$, as we saw above, giving rise to the transformations
\[L_{\alpha^{i}}(q,p)=\left\{\begin{array}{ll}
q^{'i}=q^{i}\cos(\alpha^{i})-p_{i}\sin(\alpha^{i}),\\
\\
q^{'k}=q^{k},\\
\\
p'_{i}=q^{i}\sin(\alpha^{i})+p_{i}\cos(\alpha^{i}),\\
\\
p'_{k}=p_{k}.
\end{array}\right.\]
\noindent These transformations are rotations by an angle $\alpha^{i}$ in the plane $\textup{span}(\frac{\partial}{\partial q^{i}},\frac{\partial}{ \partial p_{i}})$. Similarly, we can compose all transformations $L_{i}$ to obtain a continuous total Legendre transform:
\begin{equation}
\label{eq:L}
L_{(\alpha^{1},...,\alpha^{n})}\equiv L_{\alpha^{n}}\circ L_{\alpha^{n-1}}\circ...\circ L_{\alpha^{1}}.
\end{equation}

The set $\{L_{\alpha}\}_{\alpha \in T^{n}}$, together with the composition operation, generates an $n$-parametric subgroup of symplectomorphisms, which we can identify with the maximal torus of the group of rotations in $\mathbb{R}^{2n}$, that is, the maximum abelian subgroup of $\textup{SO}(2n)$ \cite{Adams}. Indeed, for each $\alpha \in T^{n}$, $L_{\alpha}$ generates $n$ simultaneous but independent rotations in the planes $\textup{span}(\frac{\partial }{\partial q^{1}},\frac{\partial}{\partial p_{1}})$, $\textup{span}(\frac{\partial}{\partial q^{2}} ,\frac{\partial}{\partial p_{2}})$, ..., $\textup{span}(\frac{\partial}{\partial q^{n}},\frac{\partial }{\partial p_{n}})$, by $\alpha^{1}$, $\alpha^{2}$,..., $\alpha^{n}$, respectively.

In a more general sense, we can define the left action $L:T^{n}\times T^{*}\mathcal{E}\rightarrow T^{*}\mathcal{E}$ just like $L_{\alpha}:T^{*}\mathcal{E}\rightarrow T^{*}\mathcal{E}$, thus getting a symplectic action. Regarding the Lie algebra $\mathfrak{t}^{n}\simeq \mathbb{R}^{n}$, this also has an action on $T^{*}\mathcal{E}$ which is inherited from $L$ from its diferential version, i.e., each element $\beta=\beta^{i}\frac{\partial}{\partial \alpha^{i}}\in \mathfrak{t}^{n}$ can be associated with a vector field on $ T^{*}\mathcal{E}$ defined as
\begin{equation}
\label{eq:fun_vec}
\hat{\beta}(q,p)=\begin{array}{c|}\frac{d}{ds}L_{s\beta}(q,p)\end{array}_{\,\,s=0}=\displaystyle\sum\limits_{i=1}^{n}\beta^{i}\left(q^{i}\displaystyle\frac{\partial}{\partial p_{i}}-p_{i}\displaystyle\frac{\partial}{\partial q^{i}}\right).
\end{equation}

Being $L$ a symplectic action, we can say that, at least locally, the fields (\ref{eq:fun_vec}) are Hamiltonian fields, since $0=\mathscr{L}_{\hat{ \beta} }\Omega=d(i_{\hat{\beta}\Omega})=0$ implies that for every $\beta\in \mathfrak{t}^{n}$ there exists $\mu_{\beta }\in \mathcal{C}^{\infty}(T^{*}\mathcal{E})$ such that $\hat{\beta}\in \mathfrak{X}(T^{*}\mathcal {E} )$ is its Hamiltonian vector field:
\[d\mu_{\beta}=-i_{\hat{\beta}}\Omega.\]
\noindent An important characteristic of the fields (\ref{eq:fun_vec}) is that they are linear in the Lie algebra, since in essence they are the pushforwards of the action $L$ in the identity of $ T^{n}$ \cite{MomMap}. This allows us to conclude that the assignment of the Hamiltonians $\mu_{\beta}$ is also linear in the algebra, i.e., there exists a mapping $\mu:T^{*}\mathcal{E}\rightarrow (\mathfrak{t}^{n})^{*}$, called the momentum map \cite{MomMap} such that $\langle \mu ,\beta\rangle=\mu_{\beta}$. In our particular case, we can see that this momentum map is given by
\begin{equation}
\label{eq:Momentum}
\mu(q,p)=-\displaystyle\frac{1}{2}\displaystyle\sum\limits_{i=1}^{n}\left[(q^{i})^{2}+(p_{i})^{2}\right]d\alpha^{i}.
\end{equation} 

\noindent Indeed,
\[\begin{array}{ll}
d(\langle \mu(q,p),\beta\rangle)+i_{\hat{\beta}}\Omega&=d\left(-\displaystyle\frac{1}{2}\displaystyle\sum\limits_{i=1}^{n}(dp_{i}\wedge dq^{i})\right)+\\
\\
&\phantom{=}\Omega\left(\sum_{j=1}^{n}\beta^{j}\left(q^{j}\displaystyle\frac{\partial}{\partial p_{j}}-p_{j}\displaystyle\frac{\partial}{\partial q^{j}}\right)\right)\\
\\
&=-\displaystyle\sum\limits_{i=1}^{n}\beta^{i}\left[q^{i} dq^{i}+p_{i}dp_{i} \right]\\
\\
&\phantom{=}
+\displaystyle\sum\limits_{i=1}^{n} [\beta^{i}q^{i}dq^{i}+p_{i}dp_{i}]=0.
\end{array}\]

An important feature of the momentum map (\ref{eq:Momentum}) is that it is equivariant, i.e., it maps the entire orbit $\mathcal{O}_{(q,p)}=\{L_{\alpha}(q,p)\}_{\alpha\in T^{n}}$ to an only point $\mu(q,p)\in \mathfrak{t}^{n}$,
\[\begin{array}{ll}
\mu\circ L_{\alpha}(q,p)&=\mu\left(q\cos(\alpha)-p\sin(\alpha),q\sin(\alpha)+p\cos(\alpha)\right)\\
\\
&=-\displaystyle\frac{1}{2}\displaystyle\sum\limits_{i=1}^{n}\left[(q^{i}\cos(\alpha_{i})-p_{i}\sin(\alpha^{i}))^{2}+\right.\\
\\
&\phantom{=-\frac{1}{2}\displaystyle\sum\limits_{i=1}^{n}}\left.(q^{i}\sin(\alpha^{i})+p_{i}\cos(\alpha^{i}))^{2}\right]d\alpha^{i}\\
\\
&=-\displaystyle\frac{1}{2}\displaystyle\sum\limits_{i=1}^{n}\left[(q^{i})^{2}+(p_{i})^{2}\right]d\alpha^{i}=\mu(q,p).
\end{array}\]
\noindent This means that all states in the orbit $\mathcal{O}_{(q,p)}$, which are geometrically equivalent under the generalized notion of Legendre transform that we have introduced, are characterized by the constant $n$ quantities
\begin{equation}
\label{eq:MomComp}
\mu_{i}=-\frac{1}{2}[(q^{i})^{2}+(p_{i})^{2}].
\end{equation}
\section{Conclusions}
\label{sec:con}

An important element in a geometrization program of thermodynamics is the contact structure \cite{Mrugala,Arkady,Quevedo}. From this, the symplectic nature of thermodynamics appears naturally when identifying the phase space as a contactization of the cotangent bundle of the  equilibrium space, which also has an organic construction attached to the ideas of non-geometric thermodynamics. In this sense, the Hamiltonian description of thermodynamic processes is a natural consequence of the description of these processes via contact Hamiltonians  \cite{Mrugala,Mrugala2,Bravetti}. This is at the same conceptual level, i.e., 
it is a translation of contact geometry into symplectic geometry. Likewise, the maps that relate one thermodynamic system to another are also part of the phase space with its contact structure \cite{Jacek}.

The introduction of the symplectic approach in thermodynamics is interesting for two reasons.
The first one has to do with the  simplification of the structures involved in the description of thermodynamics. Although contact geometry is an increasingly used tool in physics \cite{Arkady}, symplectic geometry offers a simple tool to explore physical systems, for example, following a canonical quantization approach, which is on the border between classical and quantum thermodynamics. As a second aspect to note is that the symplectic perspective gives us a Hamilton-Jacobi-like equation (\ref{eq:24}) whose general solution is a fundamental equation of an $n$-parametric family of thermodynamic systems, which in principle would allow us to characterize complete families of thermodynamic systems.
However, for an {\it a priori} given system we should be able to  choose among all the Hamiltonians the one that captures the physical behavior of that system.  
This problem does not seem to be easy to solve because of the large number of available Hamiltonians and thermodynamic systems. Nevertheless, it seems reasonable to conjecture that for any system it is always possible to find a Hamiltonian from which the corresponding fundamental equation can be derived.  Another indication of this freedom is the case of the Hamiltonian $h_{3}$ taken as an example of a map that relates different thermodynamic systems. This Hamiltonian flow might be locally interpreted as a generator of transformations that relate different thermodynamic systems. This resembles the Kustaanheimo-Stiefel transformation in classical mechanics that relates a two-body gravitational system with  a harmonic oscillator \cite{ks65,saha09}. It would be interesting to further investigate this similarity.

In the geometric description of thermodynamics, metric structures have also been introduced and investigated at the level of the phase space and the equilibrium space \cite{Quevedo,Quevedo2,Weinhold,Ruppeiner,Sivak}. In the case of symplectic geometry, there are different manners to introduce metric structures that could be applied to study symplectic thermodynamics as formulated in this work. This is a task that we expect to address in  subsequent works.

\begin{acknowledgements}
This work was partially supported  by UNAM-DGAPA-PAPIIT, Grant No. 114520, and 
Conacyt-Mexico, Grant No. A1-S-31269.
\end{acknowledgements}

\appendix

\section{Non-uniqueness of the contactization procedure}

The key to the contactization procedure on a manifold $\mathcal{M}$ is that given an exact symplectic structure, it is always possible to introduce it in a  co-dimension one environment, becoming a pre-symplectic structure, which naturally defines a maximum non-integrable and co-orientable contact distribution. An important detail that stands out in this context is that the distributions are generated by the kernel of the \textit{symplectic potential} $\theta$ and not directly by the symplectic form, $\Omega=d\theta$. This means  that we can associate with $\mathcal{M}$ not only a single contact manifold $\mathcal{M}\times \mathbb{R}$, but also one for each choice of the 1-form $\theta$ in the class of the De Rham's cohomology of symplectic potentials. Since $\theta$ and $\theta+dF$ are 1-forms that generate the same symplectic form, $\Omega=d\theta=d(\theta+dF)$, both 1-forms generate different contact distributions in $\mathcal{M}\times\mathbb{R}$: $\alpha_{C}\,=\,dz-\pi_{C}^{*}\theta\,\neq\,dz-\pi_{C}^{*}\theta-d\left(\pi_{C}^{*}F\right)\,=\,\alpha'_{C}$. 

At a deeper level the root of the problem comes from the fact that, when choosing a portion of a tangent space (a distribution), the complementary distribution is not unique; in this case, one chooses the vertical part and has the freedom to define the horizontal part of the distribution. On the other hand, this ambiguity is not so serious in the geometric sense, since we can always define a mapping $\Phi_{F}$ of the contact manifold with 1-form $\alpha'_{C}$ to the contact manifold with 1-form $\alpha_{C}$, such that it relates only  contact distributions: 
\begin{equation}
\label{eq:A_1}
\Phi_{F}:\,(z,x)\,\mapsto\,(z'=z+\pi_{C}^{*}F,x'=x).
\end{equation}

In effect, (\ref{eq:A_1}) is basically a vertical translation on the fibers; therefore, $\pi_{C}\circ \Phi_{F}=\pi_{C}$, this being the key of mapping a contact 1-form to a contact 1-form:
\begin{equation}
\label{eq:47}
\begin{array}{l}
\Phi^{*}_{F}\alpha'_{C}\,=\,\phi_{F}^{*}dz'\,-\Phi_{F}^{*}\circ\pi_{C}^{*}\left(\theta+dF\right)\\
\\
\phantom{\Phi^{*}_{F}\alpha'_{C}}=d(z+\pi_{C}^{*}F)-\pi_{C}^{*}\theta-d(\pi_{C}^{*}F)\\
\\
\phantom{\Phi^{*}_{F}\alpha'_{C}}=dz-\pi_{C}^{*}\theta=\alpha_{C}.
\end{array}
\end{equation}

Thus, for practical purposes, the choice of contactization is the choice of a coordinate system; therefore, without losing generality we can speak of a single contactization associated with $\mathcal{M}$.



\end{document}